# Electrostatic potential measurement at the Pt/TiO$_2$ interface using electron holography

Hiroshi Nakajima[1], Toshiaki Tanigaki[2], Takaaki Toriyama[3], Mahito Yamamoto[4], Hidekazu Tanaka[4], and Yasukazu Murakami[1, 3]

[1] *Department of Applied Quantum Physics and Nuclear Engineering, Kyushu University, Fukuoka 819-0395, Japan.*

[2] *Research and Development Group, Hitachi Ltd., Hatoyama 350-0395, Japan*

[3] *The Ultramicroscopy Research Center, Kyushu University, Fukuoka, 819-0395, Japan.*

[4] *Institute of Scientific and Industrial Research, Osaka University, Ibaraki, Osaka, 567-0047, Japan.*

*Authors to whom correspondence should be addressed: nakajima@mtr.osakafu-u.ac.jp*

**ABSTRACT**

The interface of Pt/TiO$_2$ plays an essential role in device engineering and chemical reactions. Here, we report the electrostatic potential distribution of a Pt/TiO$_2$ interface by electron holography. The decrease of the electrostatic potential exists at TiO$_2$ in the vicinity of the interface, indicating the presence of negative charge due to electron transfer from TiO$_2$ and Pt. The decrease of the electrostatic potential can be understood in the difference in work functions between Pt and TiO$_2$. This study reveals the interplay between Pt and TiO$_2$ and the usefulness of electron holography for probing the potential in nanoscale interfaces.

## 1. INTRODUCTION

Understanding the electrostatic potential at the metal/oxide interface is of vital importance for materials science and engineering. A difference in the work function between metal and oxide allows for the charge transfer, which results in the electric polarization at the interface or surface of the system. Another essential phenomenon due to charge transfer is bending of the conduction/valence band (*i.e.,* band bending) accompanied by a depression layer in the neighborhood of the interface. From an engineering point of view, band bending plays an important role in the emergence of materials functions, as it has been established in semiconductor technologies.[1,2] The band bending is also crucial for the studies of catalyst systems because of the potential role (*via* formation of a charged interface or surface) in providing an activation site for various chemical reactions.[3] Electron holography observations on the electrostatic potential can be a powerful tool for the thorough examinations of band bending, which results in the electric polarization and the depression layer (space charge layer) produced in the interface region. This point motivated us to carry out the electron holography study using a model catalyst specimen of a $Pt/TiO_2$ film.

Concerning the observations of electrostatic potential near the interface, one of the most successful applications of electron holography is of *p-n* junction produced in semiconductors. Following the early works by Frabboni et al.[4] and Rau et al.[5], electron holography has been applied to the thorough examinations about *p-n* junctions that are fabricated in several systems[6–12], and consequently provided rich information for the engineering of semiconductor devices. The electron holography was also applied to the detection of 2-dimensional electron gas accumulated at the heterogeneous interfaces, which have been attracted considerable attention from researchers on the viewpoints of both applied physics and device engineering.[13–17] Despite these intensive studies for the semiconductors and the multiple layers comprised of strongly correlated systems (*i.e.,* perovskite-type transition metal oxides), observations from catalyst systems remain yet to be a challenge because many of those specimens are in the form of nanoparticles. The complex irregular shape and the small volume of nanoparticles make the comprehensive study of the electrostatic potential difficult. For the studies of catalyst systems, an effective approach is to use a model specimen in the form of a thin film, which provides a flat smooth interface between a metal and an oxide.

The purpose of this electron holography study is to reveal the electrostatic potential distribution at the interface of Pt/TiO$_2$, which is a well-known catalyst system. Note that Pt/TiO$_2$ has catalytic actions when Pt nanoparticles are dispersed on the TiO$_2$ support, which has been applied to photocatalytic reactions (*e.g.,* oxidation of CO) and many other applications.[18–20] The observations accordingly provide useful information for the chemical engineering and material designs of catalyst.

## 2. METHODS

A thin Pt film, the thickness of which is approximately 10 nm, was deposited on a rutile TiO$_2$-(110) substrate (supplied by Crystal Base Co., Japan) by sputtering in Ar atmosphere at room temperature. Thin-foil specimens for transmission electron microscopy (TEM) studies were prepared by using a Ga focused ion beam (FIB) system, MI4000L (Hitachi). Amorphous carbon of thickness ~2 μm was deposited for the protection of the Pt surface. The carbon-deposited area was removed from the bulk substrate. Subsequently, it was attached to a Cu grid and thinned by Ga ion beams. In the thinning process, the acceleration voltage was gradually reduced from 30 to 1 kV to decrease the thickness of a damage layer. Then, the surface damage was removed by using a low voltage Ar-ion milling, NanoMill (Fischione), with the acceleration voltage of 0.9 kV. Because of this Ar-ion milling, an electron diffraction pattern acquired from the thin-foiled TiO$_2$ layer showed only negligible rings representing the amorphous damaged layer. The Supplementary material shows the evaluation results of the fabricated thin film: Intensity profiles of a high-angle annular dark-field scanning TEM (HAADF-STEM) image show a distinct lattice pattern over the entire region of TiO$_2$. Energy dispersive x-ray spectroscopy (EDS) maps depict almost flat intensity in Ti and O *K* edges, and the ratio of Ti and O was 2.3 in the vicinity of the interface. Thus, the TiO$_2$ region remains intact after Pt deposition. Furthermore, Ar was not detected by EDS. These results demonstrate that the specimen provided a suitable situation for observing phase shifts in electron holography.

Observations of TEM, HAADF-STEM were performed by using an electron microscope, ARM200CF (JEOL Co. Ltd., Japan), with the acceleration voltage 120 kV. The relative thickness of the specimen was measured by electron energy-loss spectroscopy (EELS) using ARM200CF. Electron holograms were obtained by using another microscope, HF3300X (Hitachi Co. Ltd., Japan), which was optimized for electron holography at the acceleration voltage of 300 kV. Electron holograms were recorded by using a charge-coupled device camera, Ultrascan4000 (Gatan). The electron holography microscope was equipped with a double-biprism system, which allows for producing holograms without Fresnel fringes[21]. The phase reconstruction was carried out by using Fourier transform with digitized holograms. To improve the signal-to-noise ratio, 100 holograms were obtained and averaged after the phase reconstruction: the complex images were transformed into amplitude and phase images, and then the phase images were summed after drift correction. The fringe pitch in the electron holograms was 0.45 nm.

## 3. RESULTS AND DISCUSSION

The morphology of the thin-foil specimen was thoroughly examined before the electron holography study on the electrostatic potential. Figure 1(a) shows a TEM image of the thin-foiled Pt/TiO$_2$ specimen. The plane of the interface between Pt and TiO$_2$ was almost parallel to the incident beam direction, as mentioned later in greater detail. The diffraction contrast in Fig. 1(a) indicates that the Pt layer is in a polycrystalline state, in which the grain size is on the order of nm. The interface looks rough because of the diffraction contrast. However, the interface is almost flat (roughness ~1 nm) over the observation area ~10 nm as shown in the high-resolution images of the next section.

Two methods (HAADF-STEM and EELS) were employed for revealing the thickness variation within the field of view. Because the HAADF-STEM image is less sensitive to the diffraction condition and the intensity depends on the thickness and composition, the intensity profile of HAADF-STEM can be a measure of the thickness in the area of the same composition. Figure 2(a) provides a HAADF-STEM image acquired from a region shown in Fig. 1(a). As demonstrated in Fig. 2(b), which plots the intensity (averaged along the $y$ axis) $vs$ position in the $x$ axis, the thickness is uniform over the TiO$_2$ region. Note that the signal from the Pt layer is large while that from the C layer is only negligible because the intensity of HAADF-STEM is proportional to $Z^{\alpha}$, where $Z$ is an atomic number and $\alpha$ is a number smaller than 2.[22,23]

In the same area as that shown in Fig. 2(b), the relative thickness $\frac{t}{\lambda_{path}} = ln\left(\frac{I_t}{I_O}\right)$ was determined by EELS, where $t$, $\lambda_{path}$, $I_t$, and $I_O$ stand for the specimen thickness, mean free path for inelastic electron scattering, the integrated intensity from the whole energy-loss spectrum, and integrated intensity from the zero-loss peak, respectively[24]. The relative thickness was plotted as a function of the position ($x$ axis) as shown in Fig. 2(c). Again, the region of TiO$_2$ shows only a negligible change in the parameter $t/\lambda$. Thus, for later discussions about the phase shift, which depends on both the thickness and electrostatic potential, we can rule out the thickness change in the TiO$_2$ region. Not that the thickness in the TiO$_2$ region [shown in Fig. 2(c)] was determined to be 52 nm by the method of Malis et al.[24]: In the calculation, the value $\lambda_{path}$ for TiO$_2$ was estimated to be 104 nm for the acceleration voltage $E = 120$ kV and the collection semiangle $\beta = 9$ mrad.

We briefly mention the smoothness of the Pt/TiO$_2$ interface in the specimen. As shown in the bright-field (BF) STEM image of Fig. 2(d), in which the [1$\bar{1}$0] zone axis of TiO$_2$ is parallel to the incident electrons, the contrast of TiO$_2$ lattice is obscured in the very vicinity of the interface: refer to the region indicated by the red arrow (length of less than 1 nm). The lattice image was also obscured near the interface in a HAADF-STEM image shown in Fig. 2(e). Presumably, the Pt/TiO$_2$ interface was not perfectly planar along the incident-beam direction but meandering to some extent. At the interface, Pt and TiO$_2$ are superimposed because the images are projected along the incident beam direction. Thus, determining the position of the interface was done at the accuracy of ~ 1 nm (the red arrow) when the projection of the TEM images was concerned. As discussed in the subsequent sections, this ambiguity did not hamper the examination of the electrostatic potential because the area showing the anomaly of the phase shift was larger than the ambiguous area revealed by the STEM images. We note that, in the region where the contrast is obscure, the presence of an interlayer of a Pt-Ti alloy during the film growth cannot be discernible from the HAADF-STEM image.

Electron holography observations were performed to reveal the electrostatic potential near the Pt/TiO$_2$ interface. Electron holograms [as shown in Fig. 1(b)] were collected from the same region as that of Fig. 1(a). To suppress the undesired diffraction contrast (which can provide additional phase shift in the electron wave) in the TiO$_2$ region, the specimen was tilted along the $x$ axis. However, as demonstrated in the diffraction pattern of the inset, $hh0$-type

Bragg reflections were systematically excited so that the Pt/TiO$_2$ interface remained parallel to the incident beam direction. Figure 1(c) provides a reconstructed phase image that was obtained by Fourier transform using the electron holograms. Note that Fig. 1(c) is an averaged image over 100 reconstructed phase images collected from the same field of view. Precision in the phase measurement, which was evaluated by the standard deviation of the reconstructed phase, was ± 0.015 rad. Figure 3(a) compares the plot of the phase (red curve) with the intensity profile of the TEM image (blue curve), both of which were measured along the X1-X2 line shown in Figs. 1(a) and 1(c). The position of the Pt/TiO$_2$ interface was determined at the point indicated by the dotted line, where the slope of the intensity profile steeply changed. Consistent with the results of the thickness measurements in Fig. 2(b)–(c), the intensity profile of the TEM image showed only a negligible change throughout the TiO$_2$ region. Importantly, for the TiO$_2$ region, the phase decreases in the neighborhood of the interface, while the phase shift remains almost unchanged over the region away from the interface (over 4 nm). The horizontal line (dash-dotted line) represents the fitting of the phase over the range from 4 to 10 nm. Concerning this fitting line, the change in the phase shift $\Delta\phi$ is largest at the position of the Pt/TiO$_2$ interface. The value of $\Delta\phi$ leached 0.11 rad at the interface. The change in the phase shift of TiO$_2$ results from the change in the electrostatic potential because the phase shift is proportional to the potential.[25] Note that the phase shift increases in the Pt layer because the mean inner potential of Pt is larger than that of TiO$_2$.

For further assessments of the phase shift $\Delta\phi$, the electron holograms were acquired in different conditions of specimen tilts around $x$ axis of Fig. 1. For reference, Figs. 3(b) and 3(c) show the diffraction pattern from the TiO$_2$ region and the phase plot along the X1-X2 line, which are obtained at the tilt angle of 2.2°. The results are similar to those shown in Figs. 1 and 3(a). Figures 3(d) and 3(e) provide another set of observations collected in a distinct tilt angle of 10.2°. In this tilt condition, the Pt/TiO$_2$ interface appears still parallel to the incident electrons, because of the systematic excitations of *hh0*-type Bragg reflections. Again, the phase shift in Fig. 3(e) decreases near the Pt/TiO$_2$ interface. Both the magnitude of $\Delta\phi$ and the length ($\lambda = 4$ nm) over which the phase shift decreases are consistent with those in Fig. 3(a). Thus, regarding the source of the phase shift $\Delta\phi$, we can rule out the possibility of electron diffraction that gives rise to additional geometric phase shifts. The result of Δϕ should be due to the change in the electrostatic potential near the interface.

The observation of Δϕ should be discussed in terms of the energy band diagram. When the change in the specimen thickness can be negligible, the decrease in the phase shift (*i.e.,* Δϕ shown in Fig. 3) indicates the upward band bending in the TiO$_2$ region, as schematized in Fig. 4.[5–12] Note the decrease in the phase shift represents the increase in the electrostatic potential in the band diagram when the mean inner potential is defined as positive. The upward band bending suggests the presence of the negative charge. Intuitively, this phenomenon can be explained by the charge transfer from TiO$_2$ to Pt, which is triggered by the difference of the work functions [Fig. 4(a)]. While the work function of Pt ($W_{Pt}$) can be assumed to be 5.6 eV,[26,27] that of TiO$_2$ ($W_{TiO_2}$) ranges from 4 eV to 6 eV depending of the crystal surface and the methods of measurements[27–31]. Following the recent study by Bowker and Sharpe[32], sputtering onto the TiO$_2$ crystal induces the reduction of Ti ions, which results in carrier doping in the neighborhood of the surface and accordingly affects the work function. Actually, Onishi et al.[28] reported the value of $W_{TiO_2}$ to be ~5.3 eV for a sputtered TiO$_2$ (110) specimen. Since the TiO$_2$ substrate in our specimen was also subjected to the Ar-ion sputtering as mentioned in the method section, it is likely that the work function $W_{TiO_2}$ is smaller than $W_{Pt}$. Assuming the difference of work function $\Delta W = W_{Pt} - W_{TiO_2} > 0$ as illustrated in Fig. 4(a), electrons are transferred from TiO$_2$ to Pt to make the Fermi level ($E_F$) coincide with each other. This electron transfer induces the band bending $\Delta E$ [Fig. 4(b)]. The magnitude corresponds to $\Delta W = \Delta E$ and the depression layer exists over the length $D$. In the electron holography observations, Δϕ and λ are related to the parameters $\Delta E$ and $D$, respectively.

Using the relationship of Eq. (1), we can deduce the magnitude of $\Delta E$ (in the unit of eV) from the electron holography observation of Δϕ:

$$\phi(x) = \sigma \int V dz, \quad (1)$$

where $\phi(x)$, $\sigma$, and $V$ stand for the phase measured at the position $x$, the interaction constant ($6.53 \times 10^{-3}$ rad/Vnm for the acceleration voltage 300 kV) and the electrostatic potential.[25] The line integral should be carried out along the z axis (parallel to the incident electrons). Since the magnitude of $\Delta\phi$ was determined to be 0.11 rad in the vicinity of the Pt/TiO$_2$ interface, we obtained the value of $\Delta E = 0.32$ eV. The result indicates that the work function of TiO$_2$ is smaller than that of Pt, and the value of $W_{TiO_2}$ is ~5.28 eV if $W_{Pt}$ is 5.6 eV. This value agrees well with the result from other sputtered TiO$_2$ (110) specimens ~5.3 eV.[28]

Once the magnitude of band bending $\Delta E$ and the length of depression layer $D$ ($\approx \lambda = 4$ nm) are determined by the experiments, the carrier density $\rho$ within the depression layer can be estimated by the following relationship[27],

$$\Delta E = \frac{e\rho\lambda^2}{2\epsilon\epsilon_0}, \quad (2)$$

where $e$, $\epsilon$, and $\epsilon_0$ stand for the elementary charge, the relative dielectric constant of $10^2$ for TiO$_2$, and the dielectric constant of vacuum. Calculating Eq. (2), the value of donor density $\rho$ is estimated to be on the order $10^{26}$ /m$^3$ ($10^{20}$ /cm$^3$). This value can also be comparable to the carrier densities of TiO$_2$ crystals showing conductivity[33], supporting the evidence that the depression layer is formed by the work function difference $\Delta W = \Delta E$. Moreover, the electron holography observations did not suffer from appreciable electric charging when the specimen was illuminated by incident electrons. Besides, electron irradiation should induce positive charges as revealed by electron holography.[34,35] Furthermore, even if an interlayer of a Pt-Ti alloy discussed in the STEM images is formed, it cannot explain the decrease of the phase shift: the interlayer should give rise to the increase of the phase shift because the mean inner potential of a Pt-Ti alloy is high because of the high mean inner potential of Pt, as shown in the phase image. These reasons corroborate the presence of significant carriers (negative charge) that could be induced at the interface.

## 4. SUMMARY

This work provides nanoscale insight into the fundamental nature of a Pt/TiO$_2$ interface. Phase mapping using electron holography indicates the decrease in the electrostatic potential (band bending) accompanied by charge transfer in the vicinity of the interface. Furthermore, the decreased value (0.32 V) is comparable to the difference of Fermi levels (or work functions) between Pt and TiO$_2$, supporting the presence of negative charge due to the charge transfer. This study also demonstrates the utility of electron holography as a means of visualizing the potential distribution of metal/oxide interfaces at nanoscale dimensions.

**SUPPLEMENTARY MATERIAL**

See supplementary material for the intensity profiles of HAADF-STEM and elemental maps using EDS.

**ACKNOWLEDGMENTS**

We thank Koji Shigematsu (Kyushu University) for the help with the experiments. This study was supported in part by JST CREST (JPMJCR1664) and JSPS KAKENHI (JP18H03845).

**DATA AVAILABILITY**

The data that support the findings of this study are available from the corresponding author upon reasonable request.

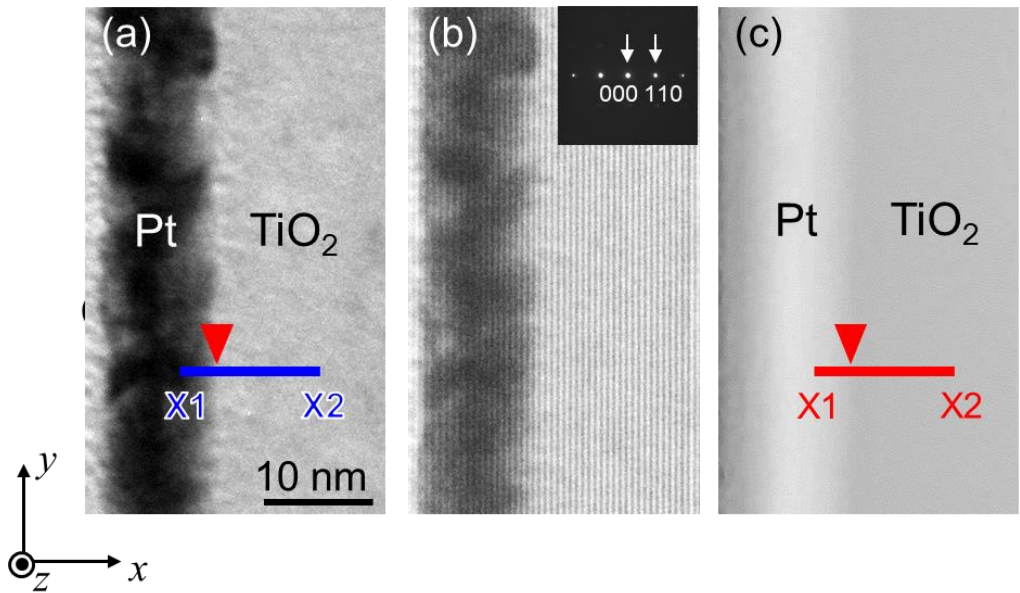

Fig. 1. Electron microscopy observation of a Pt/TiO$_2$ thin film fabricated using a focused-ion beam. (a) Bright-field image. (b) Electron hologram. Inset shows an electron diffraction pattern in the area of TiO$_2$. The specimen was tilted by 2.0° around the *x* axis. (c) Reconstructed phase image. C, Pt, TiO$_2$ represent carbon, platinum, and titanium dioxide layers, respectively. The red arrowheads represent the position of the interface.

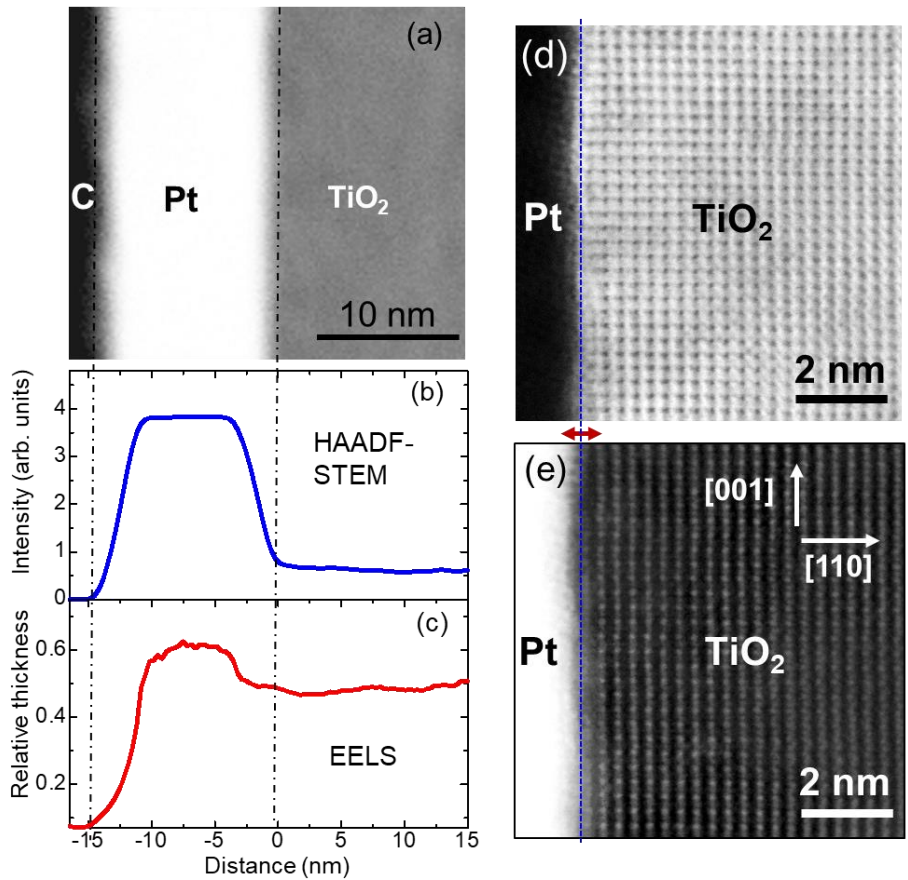

Fig. 2. Evaluation of thickness of the fabricated thin-film (a) HAADF-STEM image in the vicinity of the Pt/TiO$_2$. (b) intensity profile and (c) the relative thickness measured by EELS. The relative thickness means $t/\lambda_{path}$. The HAADF and thickness profiles were obtained simultaneously in the same position. Note that panel (b) indicates a relative thickness change in each layer because the intensity of HAADF-STEM is proportional to the thickness. (d) BF-STEM and (e) HAADF-STEM images in the vicinity of Pt/TiO$_2$. The red arrow represents the area where the contrast of the images is obscure.

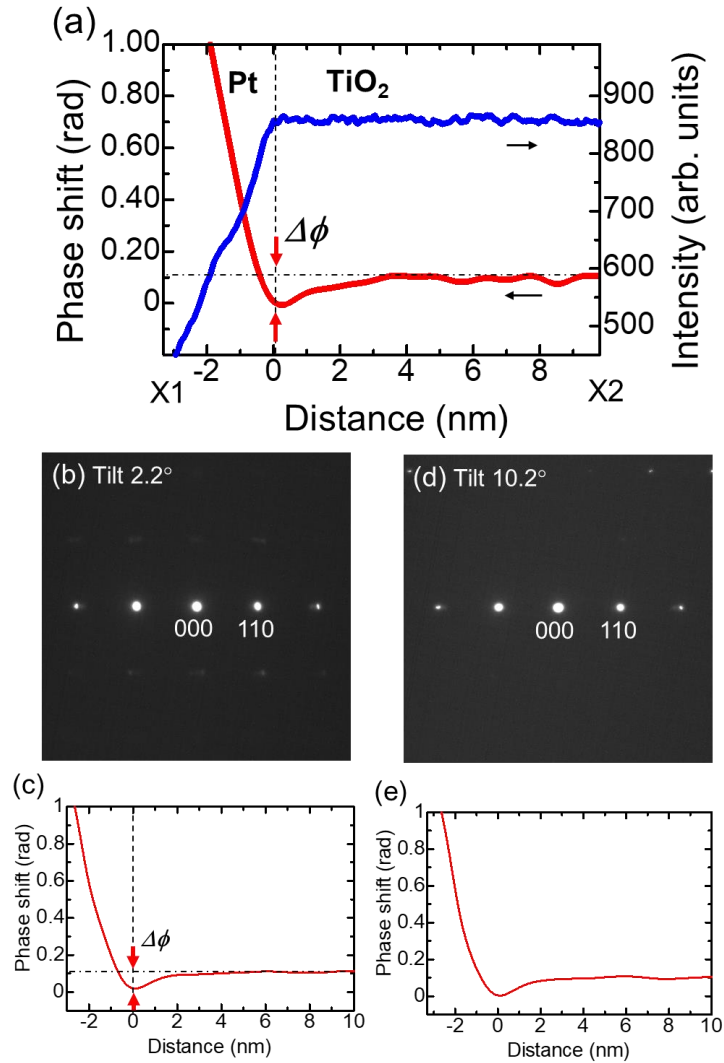

Fig. 3. (a) Intensity profiles of phase shift and bright-field images along X1–X2 of Fig. 1. The red line (left axis) represents the phase shift while the blue line (right axis) shows the intensity of TEM. Each profile is averaged over 10 nm perpendicular to the interface. The origin is defined at the interface. The dotted lines show the decrease in the phase shift in $TiO_2$ near the vicinity of the interface. The phase resolution is less than 0.02 rad. The red arrows show the decrease of the phase shift at the interface. (b) Electron diffraction patterns and (c) phase shift profile in the vicinity of the $Pt/TiO_2$ interface at the tilt of 2.2°. (d) Electron diffraction pattern at the tilt of 10.2° to investigate diffraction effects. (e) Phase shift profile at the tilt of 10.2°. No change in the phase shift was observed after the tilt.

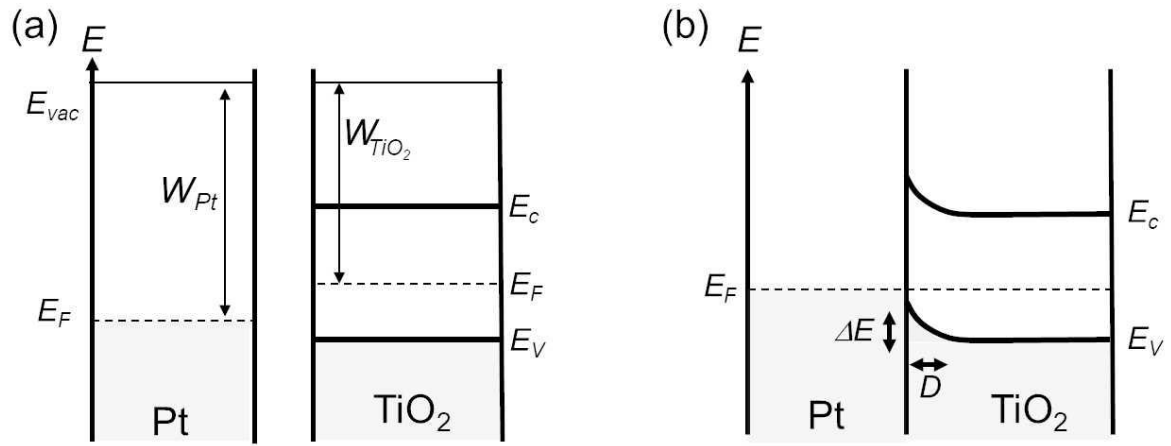

Fig. 4. Band structures of (a) before and (b) after a Pt layer is attached to TiO$_2$. $W_{Pt}$ and $W_{TiO_2}$ are the work functions of Pt and TiO$_2$, respectively. $E_F$, $E_c$, and $E_v$ represent the Fermi level, the bottom of the conductance band, the top of the valence band, respectively.